# The Intrinsic Instability of Financial Markets

Sabiou Inoua

inouasabiou@gmail.com

August 2015

**Abstract.** In this paper we explain the wild fluctuations of financial prices from the intrinsic amplifying feedback of speculative supply and demand. Formally, we show that an asset return follows a multiplicative random growth with exogenous input, which is well-known to be a generic power-law generating process, and which could thus easily explain the well-established power-law distribution of returns, and other related variables. Moreover, the theory we develop here is a general framework where competing ideas can be discussed in a unified way. The dominant random walk model, for instance, is easily derived in this framework if we superimpose market clearing (central to neoclassical economics). It corresponds to the case where the feedback in price dynamics is ignored in favor of the external input, namely the random inflow of news from the real economy. If we use instead the so-called 'price impact function', as a more realistic concept of price adjustment, then the theory leads to the multiplicative random growth model. Also, the exogenous input involves investors' average prediction error, or, alternatively, their confidence about a future change in return, so that behavioral considerations can also be addressed within the same framework.

## Contents

## 1 Introduction

**The dominant paradigm**

What's behind financial crises, notably those of 1929 and 2008? Why is financial activity so constantly fluctuating, more generally? Why are financial variables, particularly financial asset prices, so volatile compared to real economic variables?

According to the dominant paradigm, the so-called 'efficient market hypothesis', the price of a financial asset (stock, bond, currency) moves in reaction to *new information* coming to the market, which makes investors reevaluate the 'intrinsic' or 'fundamental' value of the asset, that is, the real economic perspectives of its issuer (a corporation, a state). Thus, in this view, financial markets are *externally driven*, for they are merely reacting to exogenous real news. The real sector, in other words, drives the financial one.

Formally, one assumes that financial prices move in a *random walk*, at the rhythm of the randomly coming news. This idea goes back to Louis Bachelier's pioneering thesis, *The Theory of Speculation*, in 1900—though he did not mention explicitly the underlying inflow of news, an ingredient that became central after Eugene Fama's famous review and synthesis [1-5] [1]. By convenience, however, it has become more standard to assume following a random walk, not price itself, but its logarithm[2]. That is, one assumes that $\ln P_{t+1} = \ln P_t + e_t$, where $\{e_t\}$ is a sequence of *zero-mean iid* (i.e. independent and identically distributed) *normal* variables, which represent the impacts on price of news coming to the market; more precisely, $\{e_t\}$ denotes to the constant adjustments investors make in their assessment of the security's value, in view of the news coming to them. Now, if we define return as $r_t = \ln P_{t+1} - \ln P_t$, this random walk hypothesis comes down to writing

$$r_t = e_t. \tag{1}$$

The assumption of zero-mean increments, i.e. $E(e_t) = 0$, and thus of *zero-mean return*, $E(r_t) = 0$, reflects the old view that, if speculation is to be a '*fair game*' of chance, in which a speculator is neither advantaged nor disadvantaged a priori, his expected gain should be zero (an idea colloquially phrased today by 'you can't beat the market'). As for the *independence* in the process $\{e_t\}$, and thus in $\{r_t\}$, it conveys the view that past returns are no guide for predicting future returns (so chart readers, or chartists, e.g., whose goal is to gain from hidden patterns in price data, are thus wasting their time, in this view). Finally, and most importantly, the *normality* of $\{e_t\}$, and thus of $\{r_t\}$, is often justified by the central limit theorem: if the price is to vary, news after news, by small independent increments, of *finite variance*, then monthly, weekly, or daily returns, to the extent that they are sums of large numbers of these increments, are normally distributed.

Such is the foundation of modern financial economics, underlying all matters pertaining to the allocation of financial resources, from the theory of corporate finance to portfolio theory and the pricing of complicated financial derivatives (such as options). But it has been challenged on many grounds, however; especially in view of extreme fluctuations that occurred without any special fundamental news coming to the markets—notably the crisis of October 19, 1987 [11-13]. The critique is actually vast and multidimensional. The behavioral literature, for instance, highlights the human side of financial fluctuations, which is overlooked in the dominant theory; namely, the psychological and social dimensions of investors' expectations—the fickleness of human psychology, the cognitive biases it entails, and the role of crowd psychology and herd behavior [14-19].

Particularly important, from this paper's perspective, is the critique which is based on a careful analysis of the immense wealth of high-frequency data available today, and which uncovers almost universal empirical regularities of financial fluctuations, above all, their non-Gaussian nature. This literature, initiated by Benoit Mandelbrot and involving more and more hard scientists today (especially the so-called 'econophysicists'), is a very active research program [20-22]. The distributions of returns and other financial variables are *power-law tailed*, which means that extreme events are much more likely than would allow the normal distribution. It is Mandelbrot who first observed that 'the variation of certain speculative prices', namely cotton prices, are power-law tailed with exponent $\mu \approx 1.7$, and thus have *infinite variance* [20]. What's behind this extreme randomness?

Mandelbrot's suggestion, after this discovery, is to dispense with the assumption of finite-variance increments $\{e_t\}$, which is vital to the Gaussian hypothesis, and to assume that returns follow instead a *Levy walk*. A Levy walk (or 'Levy flight') is a process of independent increments whose distribution is power-law tailed with an exponent in the range $0 < \mu < 2$, and whose variance is thus infinite. By the generalized central limit theorem, due to Paul Levy, these infinite-variance increments, when aggregated, converge, not to the normal distribution, but to one of the so-called Levy distributions, which are a special class of infinite-variance power-law tailed distributions sharing with the normal distribution the property of being stable (or invariant) under addition (i.e. a sum of Levy variables is a Levy variable, up to an affine rescaling). Power-law tailed variables, particularly those with infinite variance, exhibit an extreme randomness that Mandelbrot called 'wild', in contrast to the 'mild' Gaussian randomness [23, 24]. Shortly after Mandelbrot's work on cotton prices, Eugene Fama (then Mandelbrot's PhD student) extended it (in his dissertation) to 30 stocks of the Dow Jones, finding tail exponents below two, that is, in the Levy range [25]. Fama then sought a rationale to Mandelbrot's hypothesis in the underlying random effects of news on prices, which, if power-law tailed with infinite-variance, produce by aggregation over a month, a week, or even a day, Levy price changes. This is pure conjecture, however; it remains to explain why the price impacts of news should be infinite-variance power-law tailed.

But, actually, Mandelbrot's hypothesis doesn't stand a more thorough empirical investigation, given the huge accumulated data: stock returns, notably, are power-law tailed, indeed, but with an exponent $\mu \approx 3$ [26, 27]. Also, while a true departure mathematically, this hypothesis remains conceptually faithful to the traditional paradigm of exogenously induced price changes, which overlooks the endogenous instability of financial markets.

**The intrinsic instability hypothesis**

Following Mandelbrot's lead, authors have developed various models to account for the power-law tails of financial fluctuations. It will be too long to review them here, however (see, e.g., Lux's review [28]). Some of these models, in particular, highlight the importance of the endogenous dynamics of financial markets, namely their very functioning, the trading activity itself, as opposed to exogenous factors[3]. This paper is primarily a contribution to this alternative paradigm. We shall derive the dynamics of prices (and other related variables) from the most basic process at work in financial markets, namely, *speculative supply and demand*, which is intrinsically destabilizing (contrary to a neo-classical view). The reason is basic and can be simply highlighted. First of all, as a comparison, supply and demand for real goods is stabilizing because it creates *negative feedback* between price and quantity, due to the *law of demand*: when price rises, quantity demanded falls, making price to fall, and so on in a self-stabilizing way. Speculation, in contrast, which is but a profitable violation of the law of demand, produces *positive feedback* in price and quantity dynamics: when they *expect* an asset's price to rise, speculators buy it for a future capital gain. This is the heart of financial instability.

Indeed, the wild dynamics of financial prices comes down to the following basic amplifying mechanism: (i) when they expect a security's price to rise, traders rush to buy it; (ii) this rise in demand then raises the price effectively (by the so-called 'price impact' of demand). That is, going step by step, those traders who expect a high return buy the security, making the price to rise; others may then consider this rise as a sign that the security is worth more, or at least that its price will continue rising for a while, making them also to buy it, and thus causing the price to rise again; and so on. It doesn't matter whether the initial impulse were a fundamental news, or a purely euphoric stimulus: this is a self-fulfilling prophecy. Symmetrically, when they fear a significant price fall, some traders will rush to get rid of it, to avoid a potential loss; this rush to sell will then depress the price effectively; some may then consider this sudden price fall as signaling a bad real prospect for the security's issuer, or at least that the price will continue falling, leading them to try to get rid of it in turn; etc. In total, by this amplified urge to sell, the price will fall massively. Also, this vicious circle can equally be triggered by a bad real news or by pure panic. Thus, the extent of financial instability, in this view, has more to do with the amplifying *internal dynamics* of speculation, than the *exogenous impulses* per se (whether real news or purely psychological stimuli) [4]. These are mere triggers.

Formally, as we shall see, an asset's return evolves by this feedback mechanism as

$$r_t = a_t r_{t-\lambda} + e_t, \qquad (2)$$

where $a_t$ and $e_t$ are random variables, and the lag $\lambda = 0$ or $1$, depending on the way we shall model expectations. Now, this *multiplicative stochastic process* is one of the most convincing power-law generating processes[5], and under very general conditions, as the mathematician Harry Kesten has rigorously proven, in what has become known as *Kesten's theorem* [31]. This deep result has been extended by many authors, notably Grincevićius, Vervaat, Brandt, and Goldie [32-44]. We present it in the appendix.

Thus the old theory fails, in view of this one, because it neglects the fundamental amplifying internal dynamics of financial markets, namely, the multiplicative feedback term $a_t r_{t-\lambda}$, while focusing on the external impulse $e_t$. In reality, that feedback is behind financial fluctuations is, at least intuitively, an old and popular idea. Bachelier, for instance, conjectured (on the first page of his thesis) that 'beside fluctuations from, as it were, natural causes, artificial causes are also involved: the stock exchange acts upon itself, and its current movement depends, not only on earlier fluctuations, but also on the current state itself' [1]. In fact, as we have highlighted, this 'artificial' tendency of the financial system of acting upon itself (in a positive feedback loop), more than the 'natural causes' (from the real economy), is the main reason of its instability.[6]

In the next section (sec. 2) we shall see how this simple theory can serve as a general framework where many central issues can be addressed in a unified way. Then we will present the framework in a formal way. If we superimpose *market clearing*, as in neoclassical economics, then we are led back to the traditional paradigm, namely the random walk model (sec. 3). Thus, in this framework, the random walk is derived, rather than merely postulated. Moreover, we shall do so in a simple way, while avoiding the unrealistic assumptions and the very abstract language of the 'efficient market hypothesis', particularly its use of the controversial model of 'rational expectations', which assumes that the agents anticipate the future as if they were using a correct model of the economy![7] Now, one of Fama's other contributions to this literature is the idea that prices are good estimates of intrinsic values [3]. But, as is well-known, this needs not hold even from a 'rational expectations' framework, for there can be 'rational bubbles' [46]. It doesn't hold in this derivation as well, for there is an exponentially growing bubble in price dynamics.

If, instead of market clearing, we assume the so-called 'price impact function', as a more realistic concept of price adjustment, then the theory leads to the multiplicative random growth model (sec. 4). From this model we explain the well-known power-law regularities of return, number of trades, transaction volume, etc. Finally, we explore two other hypotheses: that (i) mimetic expectations, as described in Keynes's beauty contest metaphor, or (ii) the correlations across the assets, could explain the power-law tails of return, within the same framework, however, and still using Kesten's theorem (sec. 5 and 6).

## 2   A unifying framework

The impulse $e_t$ involves two terms essentially: (i) investors' average error in predicting the return, or, alternatively, the average size by which they think the return will change (a measure of the overall 'state of confidence' in the market); and (ii) their average assessment of the relative gap between price and 'intrinsic value'. Of course the mistakes investors make depend on the amount and quality of the information available to them, their capacity of processing it, and other psychological considerations (such as cognitive biases and emotions) that we take as given, as *exogenous*.

We have therefore in this theory a general framework where major competing ideas can be discussed in a unified mathematical way: To what extent do real news drive security prices? What about investors' cognitive limitations or other psychological dimensions (confidence, panic) that affect their judgments? We have already suggested that these exogenous factors, though important as triggers, play a minor role compared to the intrinsic amplifying power of speculation. Now we can see why, in a precise quantitative way.

First of all, the only natural way of assessing the relative importance of these factors is to determine their respective contribution to the well-established financial regularities, namely the power-law tails. In this respect, none of these exogenous factors is important per se. Indeed, by Kesten's theorem, it is the intrinsic feedback of markets, whose extent is given by $a_t$, that accounts naturally for the power law tails, regardless of the distribution of the exogenous input $e_t$, in general. (Obviously, human psychology, for instance, cannot account for the power laws by itself; for, otherwise, we would witness the same wild randomness throughout the real economy, or in every aspect of human affairs for that matter, as these are also psychologically affected.)

Mathematically, this can be easily highlighted, step by step, assuming as a starting point the nonrandom equivalent of (2). Thus consider a system that evolves as $r_t = ar_{t-1} + e_t$, starting from some $r_0 > 0$, where $a$ is a constant and $e_t$ is a nonrandom exogenous input. Only two possibilities exist for this system. If $|a|>1$, $r_t$ diverges (exponentially) to infinity, even if $e_t = 0$: the system would be called *intrinsically unstable* (for it is unstable even without any external disturbance). Such system is bound to explode (as no quantity can grow forever). Otherwise, if $|a|\leq 1$, the system can be said to be *intrinsically stable*, for it is stable as long as the external influence $\{e_t\}$ is stable. But even then (as is well-known in Keynesian theory) the system can unleash a potentially destabilizing amplifying power, which is easily highlighted if we assume in addition that $e_t = e$, namely a constant, and $|a|<1$, namely asymptotic stability, which we assume from now on. In this case, the system converges to the equilibrium state $r^* = ke$, where $k = (1-a)^{-1} = \sum_{\lambda=0}^{\infty} a^\lambda$, the so-called *multiplier*, quantifies this intrinsic amplifying power. Thus, even when the system is internally stable, it amplifies any external disturbance by $k>1$, since $\Delta r^* = k\Delta e$ (if, e.g., $a=0.9$, any external disturbance is amplified by 10). So, if indeed external disturbances play an important role as triggers, the extent of the fluctuations they inflict on the system depends on the system's internal functioning, which is given by $a$.

Now, let $\{e_t\}$ be an iid stochastic process, $a$ remaining constant (namely, a classic first order autoregressive process). Then the system's internal stability ($|a|<1$) will manifest itself by its *stationary* outcome $\{r_t\}$. Still, external disturbances tend to be amplified on average by $k$, at the stationary state, since, then, $E(r_t) = aE(r_t) + E(e_t)$; also, the output is more volatile than the input, since $\mathrm{var}(r_t) = a^2 \mathrm{var}(r_t) + \mathrm{var}(e_t)$, that is, $\mathrm{var}(r_t) = k'\mathrm{var}(e_t)$, where $k' = (1-a^2)^{-1} > 1$. But, crucially, *the system always produces a mild (normal) output from a mild input* (in particular, stationarity and finite-variance are synonymous here).

In contrast, let $a = a_t$ vary randomly also, in an iid way. Then, by Kesten's theorem, the system will eventually settle down to a stationary course if $E(|a_t|) < 1$; also, as previously, it has an amplifying power on average, only now given by $k'' = [1 - E(a_t)]^{-1}$, since at the stationary limit $E(r_t) = E(a_t)E(r_t) + E(e_t)$ (assuming $a_t$ is independent of $r_{t-1}$). But contrary to the previous case, and this is a deep property, the system is now intrinsically unstable in that *it produces a wild (power-law) outcome even from a mild (normal) input* (in particular, it can produce an infinite-variance output from a finite-variance input, while being stationary). This is the way financial instability comes about, as we shall see.

But before moving to the new approach, we derive first the old random walk model.

## 3 The random walk model

### 3.1 Basic definitions and notations

Let $Q_{it}$ denote the number of units of a financial asset (stock, bond, currency) that an investor $i$ holds at *the beginning of period* $t$. Then, by definition, $q_{it} \equiv \Delta Q_{it} \equiv Q_{it+1} - Q_{it}$ corresponds to his buying additional units of it if $q_{it} > 0$, or his selling units of it if $q_{it} < 0$, *during period* $t$. Thus the same variable, $q_{it}$, denotes both demand and supply, except that supply is *negatively counted*, which is a standard and convenient convention. Also, let $P_t$ denote the asset's price *at the beginning of period* $t$. The relative price change occurring *during period* $t$, also called *return*, is $r_t \equiv (P_{t+1} - P_t)/P_t$. Over very short time periods (high-frequency data), where returns are of quite small magnitudes, $r_t \approx \ln P_{t+1} - \ln P_t$.

Supply and demand in financial markets take different forms called (buy and sell) *order flows*. We shall consider not only *market orders*, which are requests to transact immediately at the best available price, but also *limit orders*, which are requests to transact at a given price or better [47]. Thus, if market orders tend to be executed immediately, limit orders are executed only when a counterpart is met, which meets the price limits. Also, since not every desire to buy is instantaneously matched by a symmetric desire to sell, and vice versa, a queue of orders, especially limit orders, builds up in what is known as the order book: *markets don't clear instantaneously*. The level of this queue of orders awaiting counterparts is called the order book depth or *market liquidity*, as it measures the market's capacity of absorbing new coming orders (later on we shall denote it $L_t$).

Finally, aggregate excess demand, that is, the aggregate imbalance between supply and demand, which is the fundamental variable that drives the price, is simply $q_t \equiv \sum_{i=1}^{N_t} q_{it}$, where $N_t$ denotes the number of traders that take place in the market during period $t$, or, more conveniently, the number of (buy or sell) orders that occur during this period.

### 3.2 Deriving the random walk model

**Pure speculation**

A speculator, by definition, is someone who buys when he expects a price rise, and sells when he expects a price fall. Thus, let $r_{it}^e = (P_{it+1}^e - P_t)/P_t$ denote the return investor $i$ expects from the security during period $t$ ($P^e_{it+1}$ being the price he expects to prevail at the end of $t$). Then the speculative motive for holding the asset can be written simply as

$$q_{it} = \alpha r^e_{it}, \tag{3}$$

where $\alpha > 0$. That is, when a trader expects a price rise ($r^e_{it} > 0$) he buys proportionally[8], for a future capital gain; symmetrically, when he expects a price fall ($r^e_{it} < 0$), he sells proportionally, to cushion a potential loss. The parameter $\alpha$ represents the typical quantity bought (resp. sold) by a trader who anticipates a one percent price rise (resp. fall); it measures therefore the strength of the speculative tendency in the market.

**Dealing with expectations[9]**

We define the investors' errors of prediction as

$$\varepsilon_{it} = r_t - r^e_{it}. \tag{4}$$

In any period, some investors will overestimate the future return, while others will underestimate it, so that the individual mistakes tend to cancel on average; thus we can assume

$$E(\varepsilon_{it}) = 0. \tag{5}$$

We also assume that $\text{var}(\varepsilon_{it}) = \sigma^2 < \infty$.

**Equilibrium price dynamics**

From (3) and (4) it follows that aggregate excess demand is simply

$$q_t = \alpha N_t r_t - \alpha N_t \bar{\varepsilon}_t, \tag{6}$$

where $\bar{\varepsilon}_t \equiv \sum_{i=1}^{N_t} \varepsilon_{it} / N_t$ denotes the average prediction error.

If we assume, as in neoclassical economics, that markets clear all the time (which is counterfactual), then we need only consider market clearing or 'equilibrium' returns, namely the special values $r^*_t$ for which demand is exactly and instantaneously matched by supply: $q_t(r^*_t) \equiv 0$. From (6) it follows immediately that

$$r^*_t = \bar{\varepsilon}_t. \tag{7}$$

**Implications**

By the law of large numbers, equilibrium return converges to zero as $N_t \to \infty$, as long as the errors of predictions are weakly correlated:[10]

$$r^*_t \to 0 \text{ in probability.} \tag{8}$$

Price, in other words, converges in probability to some nonrandom equilibrium value $P^*$. Thus in a competitive speculative market (with many transacting investors) or, more simply, in a very active market (where a large number of transactions take place each period), returns tend to vanish: investment in the market is a 'fair game' in a stronger and more practical sense than merely observing that $E(r^*_t) = E(\bar{\varepsilon}_t) = 0$. Thus, as long as markets clear, speculation is not destabilizing.

More precisely, by the central limit theorem,

$$r^*_t \to \sigma \nu_t / \sqrt{N_t} \text{ in distribution,} \tag{9}$$

where $\nu_t$ is a standard (i.e. a zero-man and unit-variance) Gaussian fluctuation. Thus returns are Gaussian in a competitive markets (where $N_t$ is sufficiently large). Also, assuming all observable prices are equilibrium prices (a central postulate of neoclassical economics), we have, from (9), $P^*_{t+1} = (1 + \sigma \nu_t / \sqrt{N_t}) P^*_t$, that is, $\ln P^*_{t+1} \approx \ln P^*_t + \sigma \nu_t / \sqrt{N_t}$: the price follows an 'exponential random walk'.

## 3.3 Intrinsic value and speculative bubble

The random walk model follows directly from market clearing in a speculative market, by the central limit theorem. This is a restatement of Bachelier's model. Fama and others added to this theory the proposition that any security has an intrinsic value, or fundamental value, which investors can accurately guess, using all available information. Moreover, the price is said to wander randomly about this intrinsic value [3]. Is this justified in this framework? We shall prove this idea to be incompatible with the very logic of speculation, even if investors could accurately assess the intrinsic value on average.

Let the asset's intrinsic value be $F$, and assume it is constant in the short run, without loss of generality. Let $F_{it}$ denote investor $i$'s guess of this intrinsic value, and assume

$$E(F_{it}) = F, \qquad (10)$$

that is, the guesses are accurate overall. Beside the purely speculative motive, an investor can buy or sell the asset for fundamental reasons: its real economic perspectives. Assume, as is standard in the literature, that an investor's fundamental demand or supply is proportional to the relative value-price gap $\phi_{it} = (F_{it} - P_t)/P_t$. Thus, when $i$ thinks the asset is undervalued he buys it; otherwise, he sells it. Individual demand and supply is now

$$q_{it} = \alpha r_{it}^e + \gamma \phi_{it}, \qquad (11)$$

where $\gamma > 0$ measures the strength of the fundamental motive for holding the security, namely the typical quantity of the security bought (resp. sold) by a purely fundamentals-oriented trader who thinks the security is one-percent undervalued (resp. overvalued).

**Equilibrium**

Aggregate excess demand is now

$$q_t = \alpha N_t r_t - \alpha N_t \bar{\varepsilon}_t + \gamma N_t \bar{\phi}_t, \qquad (12)$$

where $\bar{\phi}_t \equiv \sum_{i=1}^{N_t} \phi_{it}/N_t$ denotes the average perceived price-value gap. It is easy to see that $\bar{\phi}_t \equiv (\bar{F}_t - P_t)/P_t$, where $\bar{F}_t \equiv \sum_{i=1}^{N_t} F_{it}/N_t$. The market clearing return is now

$$r_t^* = \bar{\varepsilon}_t - \rho \bar{\phi}_t, \qquad (13)$$

where $\rho \equiv \gamma/\alpha$ measures the strength of the fundamental motive in terms of the speculative one.

**Implications**

From (13), $E(r_t^*) = 0$ only if $E(\bar{\phi}_t) = 0$, i.e. only if $E(\bar{F}_t/P_t^*) = 1$, which needs not hold; so, the idea of 'fair game' is lost. Also, by the law of large numbers, $\bar{\varepsilon}_t \to 0$, as previously, and $\bar{\phi}_t \to (F - P_t)/P_t$, since $\bar{F}_t \to F$. Thus, for a large $N_t$, and neglecting the tiny normal fluctuations to focus on the trend, (13) implies that $(P_{t+1}^* - P_t^*)/P_t^* = -\rho(F - P_t^*)/P_t^*$ (still assuming that only market clearing prices realize). That is,

$$P_{t+1}^* = (1+\rho)P_t^* - \rho F. \qquad (14)$$

Thus, letting $B_t = P_t^* - F$ be the bubble component in price (i.e., the deviation of price from the intrinsic value), we have, from (14),

$$B_{t+1} = (1+\rho)B_t, \qquad (15)$$

namely, an exponentially growing speculative bubble. So rather than wandering about the intrinsic value, the price diverges exponentially from it! Why is this so?

**Speculation versus the law of demand**

This price-value divergence happens by the very logic of speculation, which creates a positive feedback that amplifies any disturbance exponentially. As we said, this doesn't happen for ordinary real goods and services, by the stabilizing virtue of the law of demand. Now we can highlight the difference mathematically.

Assume that demand were only motivated by fundamental reasons. That is, let only supply be (partially) speculatively motivated. Moreover, assume that the law of demand applies. Mathematically, all this comes down in essence to making the change $\alpha \to -\alpha$, which would create negative feedback. More precisely, assume demand and supply obey

$$q_{it} = \gamma \phi_{it} - \alpha r_t. \tag{16}$$

This says that an investor's desire to buy the security for intrinsic reasons is partially offset by a price rise, making the security less affordable; or when the price rises, an investor who holds many units of the security can sell some of them for an immediate capital gain, only this speculative tendency is partially offset by his desire to hold the security for intrinsic reasons. The implication of this assumption is immediate, for, as we said, it merely implies the change $\alpha \to -\alpha$, or $\rho \to -\rho$ in (14) and (15). Then, $P_t^* \to F$, or, equivalently, $B_t \to 0$, provided only that $|1-\rho|<1$, i.e. $\rho \leq 2$, or $\gamma \leq 2\alpha$. In sum, Fama's proposition holds only when demand is non-speculative, while the classic random walk hypothesis holds only in a purely speculative market! This is a paradox[11]. Of course, the law of demand bounds the speculative mania somehow, for speculators' spending powers are limited. This budget constraint is greatly lighten, however, by the easy availability of credit. So easy credit is the other facet of the coin we are describing in this paper.

## 4 A more scientific model

### 4.1 Empirical facts

**The non-Gaussian behavior of markets**

Even if we disregard the previous paradox, the random walk model implies that speculative price changes are Gaussian. But, again, the actual fluctuations are wilder[12]: the distribution of returns, and other variables, is power-law tailed. This regularity is particularly universal in stock markets, as it holds internationally, for various stocks, and on various time scales. We should keep in mind three well-established power laws in what follows: that of absolute stock return, $|r|$, that of the number of trades $n$, and that of the number of traded shares (or volume), $v$, on various time periods [49, 50]. That is, these variables have tail distributions that are asymptotic to powers; namely, for sufficiently large $x$,

$$P(|r|>x) \sim C_r x^{-\mu_r}, \mu_r \approx 3; \tag{17}$$

$$P(n>x) \sim C_n x^{-\mu_n}, \mu_n \approx 3.4; \tag{18}$$

$$P(v>x) \sim C_v x^{-\mu_v}, \mu_v \approx 1.5; \tag{19}$$

where $C_r$, $C_n$ and $C_v$ are constants (here and throughout $\sim$ denotes asymptotic equality). These facts invalidate the (Gaussian) random walk model. But what's behind this failure?

**Markets don't clear instantaneously**

The random walk model, as formulated here, derives essentially from market clearing, which is a central axiom in neoclassical economics, but which hardly holds in real markets. Obviously not every desire to buy would instantaneously meet an exactly symmetric desire to sell, as we said[13]. The very existence of queues of orders waiting to be executed, that is waiting to find counterparts, invalidates this idea of instantaneous market clearing. So we need a more realistic and empirically grounded alternative to this mainstream postulate; that is, we need a more realistic concept of price adjustment.

Actually, an entire body of research has already addressed this issue: the so-called 'price impact' literature. Price impact refers to the basic idea that price adjusts in reaction to a supply and demand imbalance: buy orders tend to rise the price while sell orders tend to lower it; in other words, a positive excess demand makes price to rise, and a negative excess demand makes it to fall. This idea is formalized by the so-called *price impact function*, which gives the typical price change caused by a given level of excess demand; that is, $E(r_t | q_t)$. Different time scales, levels of aggregation, and other considerations, give different price impact functions. When averaged over many transactions, however, a *linear* function fits the data very well[14]. Indeed, as Plerou et al. show, the impact of a single transaction $q_i$, or $E(r_t | q_i)$, behaves like a hyperbolic tangent, that is, almost linearly for small $q_i$, while saturating for bigger $q_i$, whereas a clear linear pattern emerges by aggregating over many transactions [52]. This aggregate linearity seems to be a general result [53]. Also, as Cont et al. point out recently, the price impact is clearly linear if we include all order flows. Indeed much of the price impact literature is about market orders only, thus ignoring limit orders, which are known to play an equally important role in price movements [54]. A better measure of supply-demand imbalance, as they show, should include all intentions to transact, as expressed by all orders (whether executed, or waiting to be executed). With this broader measure, price impact is simply[15]

$$E(r_t | q_t, L_t) = \beta q_t / L_t, \tag{20}$$

where $L_t$ stands for liquidity (or order book depth), which is crucial to condition on, and which the authors measures by the average queues of buy and sell orders during period $t$; $\beta$ is a positive constant below unity. This is both a simple, deep and intuitive relation: price reacts to a supply-demand imbalance, but the higher the market's capacity of absorbing this imbalance, through past pending orders, the smaller the impact will be.

From (20) it follows that $r_t = \beta q_t / L_t + \delta_t$, where $\delta_t$ captures tiny disturbances, such as rounding errors (coming e.g. from considering price as changing continuously, while in reality it changes only by tiny discrete units called 'ticks'). But throughout we let $\delta_t = 0$, as this noise wouldn't add anything deep to the discussion; thus we write simply

$$r_t = \beta q_t / L_t. \tag{21}$$

There is another proposition in Cont et al.'s paper that we shall use later: that excess demand behaves like the square root of volume, under reasonable conditions, by the central limit theorem; that is, as $N_t \to \infty$,

$$q_t \to c\nu_t \sqrt{v_t} \text{ in distribution}, \tag{22}$$

where $c$ is a constant and $\nu_t$ is a standard normal variable. This statistical relationship would be, according to the authors, the explanation of the so-called 'square root law' of impact, namely that price varies as the square root of volume, which is a more popular formulation of the price impact function[16].

## 4.2 Explaining the empirical facts

**Feedback in price dynamics**

As announced in the introduction, price fluctuations are amplified by the circular causation between price and excess demand, as expressed by (12) and (21). To avoid an arbitrary minus sign[17], we redefine the prediction error as $\varepsilon_{it} = r_t^e - r_t$. Thus,

$$q_t = \alpha N_t r_t + \gamma N_t \overline{\phi}_t + \alpha N_t \overline{\varepsilon}_t,$$
$$r_t = \beta q_t / L_t. \tag{23}$$

It follows the following feedback in price dynamics (putting the first eq. into the second):

$$r_t = a_t r_t + e_t, \tag{24}$$

where

$$a_t = \alpha \beta N_t / L_t, \ e_t = a_t(\overline{\varepsilon}_t + \rho \overline{\phi}_t), \tag{25}$$

and, again, $\rho = \gamma / \alpha$. The term $(\overline{\varepsilon}_t + \rho \overline{\phi}_t)$ reflects investors' abilities of prediction (of future returns and fundamental values), that is, their capacity of processing information, and other psychological aspects that affect their judgment. We shall refer to it as the *behavioral impulse* or $b$:

$$b_t = (\overline{\varepsilon}_t + \rho \overline{\phi}_t). \tag{26}$$

Now, the external influence in price dynamics terms out to be multiplicative itself; it is the product of a behavioral quantity, namely $b$, and an objective quantity, namely $a$:

$$e_t = a_t b_t. \tag{27}$$

Price changes result from the amplifications of this multiplicative input, as (24) implies:

$$r_t = k_t a_t b_t, \tag{28}$$

where the multiplier $k_t = (1 - a_t)^{-1}$.

**Explaining the power-law tails of return**

This feedback explain the power-law tails of price fluctuations, by Kesten's theorem. This theorem applies more generally when the feedback of $r$ on itself happens after one period lag, as is more common; but we shall come to this point later. The precise conditions of Kesten's theorem are listed in the appendix (theorem 1 and 2). But we can derive most of them intuitively, and show why a power law emerges naturally from this feedback.

Let $G$ be the tail distribution function of absolute return, that is, $G(x) = P(|r| > x)$, for $x > 0$. Throughout, we assume that the sequence $\{(a_t, e_t)\}$ is iid. As a starting point, we can assume a purely speculative market, or $\gamma = 0$. Then, omitting the time subscript, it follows from (24) that $P(|r| > x) = P(|ar + e| > x) = P(|ar + a\overline{\varepsilon}| > x) = P(|r + \overline{\varepsilon}| > x/a)$. By the law of large numbers, $\overline{\varepsilon} \to 0$ as $N \to \infty$; hence $|r + \overline{\varepsilon}| \to |r|$, which implies that $P(|r + \overline{\varepsilon}| > x/a) \to P(|r| > x/a)$, assuming $G$ is continuous. Therefore, for a sufficiently large $N$, we have $P(|r| > x) = P(|r| > x/a) = \int P(|r| > x/A) f(A) dA$, where $f$ stands for the density function of $a$. Therefore, $G$ obeys the functional equation

$$G(x) = \int G(x/A) f(A) dA. \tag{29}$$

It is easy to see that none of the usual distributions (normal, lognormal, exponential, etc.) satisfies this equation, except the power law

$$G(x) = Cx^{-\mu}, \tag{30}$$

which solves it naturally as follows: $Cx^{-\mu} = \int C(x/A)^{-\mu} f(A) \mathrm{d}A = Cx^{-\mu} \int A^\mu f(A) \mathrm{d}A = Cx^{-\mu} E(a^\mu)$. That is, $G(x) = Cx^{-\mu}$ is a solution of (29) as long as

$$E(a^\mu) = 1, \qquad (31)$$

or $(\alpha\beta)^\mu E[(N/L)^\mu] = 1$. This is the fundamental condition of Kesten's theorem (see appendix, theorem 1, condition (i)). It says that the tail exponent of return (which is the key statistical parameter, as it characterizes the whole distribution, up to a normalizing constant $C$) doesn't depend on the distribution of $\bar{\varepsilon}$, or $e$, more generally, as we shall see shortly, but is instead given by that of $a$. That is, the power law of return is entirely given by the feedback term $a$ (independently of the input $e$), as announced earlier.

In the general case where trading can be both speculative and fundamentals-motivated ($\gamma \neq 0$), $P(|r| > x) = P(|ar + ab| > x) = P(a|r||1 + b/r| > x) = P(|r| > (x/a)/|1 + b/r|)$. Now, when $|r| \to \infty$ (i.e. $r \to \pm\infty$), $|1 + b/r| \to 1$, *assuming that $b$ fluctuates relatively mildly, i.e. it has rarer extreme realizations than return*. This is precisely the case here since $b$ is normal by the central limit theorem (but more on that later). Also, when $N \to \infty$, $\bar{\varepsilon} \to 0$ and $\bar{\phi}$ is arguably small, making for a small $b$, which is one more reason to assume $|1 + b/r| \to 1$. In sum, when $|r| \to \infty$, and a fortiori if in addition $N \to \infty$, we can write $G(x) \sim P(|r| > x/a) = \int G(x/A) f(A) \mathrm{d}A$, which leads essentially to the same conclusion as previously, namely a power law solution; only here we are conditioning on large realizations of $|r|$. In sum, any function $G$ obeying $G(x) \sim Cx^{-\mu}$ when $x \to \infty$ will do, provided as previously, that $E(a^\mu) = 1$.

We have only shown, however (and purely intuitively), that a power-law tail is a plausible solution. This heuristic is conclusive only if the functional equation has a unique solution. But precisely, by Kesten's theorem, if $E[\ln(a)] < 0$, $r$ has a unique distribution, which is stationary. This condition is satisfied when $E(a) < 1$, since, by Jensen's inequality, $E[\ln(a)] < \ln E(a)$. Does $a = \alpha\beta N/L$ meet this condition? First, $\alpha N$ is but the total demand (resp. supply) that investors make when they anticipate a one-percent price rise (resp. fall); thus we can assume $\alpha N < L$ on average (the flow of coming orders is small compared to the stock of accumulated pending orders). In sum, therefore, we can indeed assume that $E(a) = \beta E(\alpha N/L) < 1$ (since $\beta < 1$).

Moreover, the distribution of $r$ is *necessarily power-law*, by Kesten's theorem, provided that the distributions of $a$ and $e$ obey few light conditions, of which (31) is the most important. The other conditions are trivially met here, in particular because $b$ is normal.

Empirically, $\mu_r = 3$; from this we can explain the other power laws, i.e. (18) and (19).

**Explaining the other power laws**

First of all, if $a$ is power-law with exponent $\mu_a$, then (31) requires that $\mu_a < \mu$, for otherwise $E(a^\mu) = \infty$; in general, for (31) to make sense, $a$ should be milder than a power law with exponent $\mu$. Now, if the total number of orders $N$ has the same distribution as the number of trades $n$, which we know is power-law with exponent $\mu_n \approx 3.4$, then $a$, namely $\alpha\beta N/L$, would also be power-law with exponent $\mu_a = \mu_n \approx 3.4$ (assuming that $L^{-1}$ is mild enough)[18]. Then the model would be falsified were $\mu_r \geq \mu_n$ (for then $E(a^{\mu_r}) = \infty$), which is not the case empirically, adding to the plausibility of this model.

Excess demand follows essentially the same dynamics as price:

$$q_t = a_t q_t + e'_t, \qquad (32)$$

where $e_t = \alpha N_t b_t$. It follows that return and excess demand have the same tail distributions (which is already implied by the price impact equation); that is,

$$P(q > x) \sim C_q x^{-\mu_q}, \text{ where } \mu_q = \mu = \mu_r \approx 3, \text{ and } C_q \text{ is a constant.} \tag{33}$$

Finally, from the statistical relationship (22), namely $q \to c\nu v^{1/2}$, it follows that volume is also power-law with exponent $\mu_v = \mu_q / 2 = 3/2$, as found empirically.

### 4.3 An alternative formulation

**Expectations reconsidered: investors' confidence**

So far, in dealing with expectations (of return and intrinsic value) we have focused on investors' errors of predictions, and assumed they cancel each other on average, *as long as the individual errors are weakly dependent*. To some extent, however, we can dispense with the assumption of zero expected mistake, *as long as the resulting average prediction error is milder than the feedback term a*, because then it wouldn't affect the distribution of returns, which is determined solely by $a = \alpha \beta N / L$ (rather than $b = \bar{\varepsilon} + \rho \bar{\phi}$) [19].

But we can model expectations alternatively, in a way that reflects a more fundamental driving force of investors' behavior: *confidence*. Confidence in future price change is obviously the key determinant in a speculator's decision to buy or sell. Thus, were we interested in price expectation, we would define $\theta_{it} = P_{it}^e - P_{t-1}$, so that an optimistic or 'bullish' investor (one who expects a higher price than it is today) is characterized by $\theta_{it} > 0$, and a pessimistic or 'bearish' investor by $\theta_{it} < 0$. A 'bullish market' would be that where $E(\theta_{it}) > 0$, and a 'bearish market' that where $E(\theta_{it}) < 0$. But, following Bachelier, we can assume $E(\theta_{it}) = 0$, given that 'the market, that is to say, the totality of speculators, must believe, at a given instant, neither in a price rise nor in a price fall, since, for each quoted price, there are as many buyers as sellers'; in other words, we can assume a market where bullish and bearish tendencies cancel one another.

Bachelier's justification is that buyers and sellers in a purely speculative market must have opposite views, if transactions are to take place: a buyer should think the price will rise and the seller should think the price will fall. These opposite views thus tend to cancel each other (and when most views are anti-correlated, the law of large numbers applies even more strongly than if they were independent). So we can also partially dispense with the assumption of strictly independent expectations; but only partially, as we shall see.

Only, the important variable for the investor, is not price itself, but return; so we apply the same idea directly on return and assume

$$r_{it}^e = r_{t-1} + \theta_{it}, \tag{34}$$

$\theta_{it}$ reflecting $i$'s confidence about a higher future return ($\theta_{it} > 0$) or a lower one ($\theta_{it} < 0$).

**Bachelier anticipating Keynes**

We are using these simple and general formulations in this paper to avoid unrealistic and unnecessary assumptions about investors' abilities of guessing the future. In particular, here we are merely assuming, as Keynes did in *The General Theory*, that an investor anticipates the future (return) taking the present (return) as a reference, to which we add a quantity to reflect his belief in a future change ('our usual practice being to take the existing situation and to project it into the future, modified only to the extent that we have more or less definite reasons for expecting a change', as Keynes put it [57, chap. 12]).

Also, because, realistically, confidence is more driven by 'animal spirits' (or 'spontaneous optimism' and pessimism) than the result of cold calculations, we cannot reduce it to a rigid mathematical rule; that is, we should take the additive component $\theta_i$ as given, as an exogenous random variable, and consider only its distribution across individuals (for there is 'not much [more] to be said about the state of confidence *a priori*. Our conclusions must mainly depend upon the actual observation of markets and business psychology.'). In any case, 'human decisions affecting the future, whether personal or political or economic, cannot depend on strict mathematical expectation, since the basis for making such calculations does not exist; and that it is our innate urge to activity which makes the wheels go round, our rational selves choosing between the alternatives as best we are able, calculating where we can, but often falling back for our motive on whim or sentiment or chance'.

This absence of rigid regularity in individual expectations does not prevent a precise collective pattern to emerge, of course, which can be derived mathematically (by the law of large numbers or the central limit theorem). One thing, in other words, is to assume people guessing the future by computing mathematical expectations, and another is to consider the mathematical expectations of their unmathematical guesses.

But long before Keynes, Bachelier had already observed that

*Two kinds of probabilities can be considered:*
*(1) Probability that might be called mathematical; that which can be determined a priori; that which is studied in games of chance.*
*(2) Probability depending on future events and, as a consequence, impossible to predict in a mathematical way.*
*It is the latter probability that a speculator seeks to predict. He analyses the reasons which may influence rises or falls in prices and the amplitude of price movements. His conclusions are completely personal, since his counterparty necessarily has the opposite opinion.*

**The model restated**

With this alternative representation of expectations, the dynamics of return becomes

$$r_t = a_t r_{t-1} + e_t, \qquad (35)$$

where $a_t = \alpha\beta N_t/L_t$, $e_t = a_t(\rho\bar{\phi}_t + \bar{\theta}_t)$, $\bar{\theta}_t$ being the average amount by which investors think the return will rise or fall (a summary of the 'state of confidence' in the market):

$$\bar{\theta}_t \equiv \sum_{i=1}^{N_t} \theta_{it} / N_t. \qquad (36)$$

Here the behavioral impulse is

$$b_t = \bar{\theta}_t + \rho\bar{\phi}_t. \qquad (37)$$

Compared to the previous formulation, where price feedback was instantaneous, this stochastic recurrence equation has the advantage of highlighting the dynamic nature of the feedback. Mathematically, however, the two formulations are intimately related: the first can be seen as the steady state of the second. More precisely, under the light conditions of Kesten's theorem, and no matter how initiated, the process $\{r_t\}$ given by the recurrence (35), where $(a_t, b_t)$ are iid (or at least stationary and ergodic) copies of some random pair $(a,b)$, converges in distribution to the unique stationary solution $r$ of the stochastic equation $x \stackrel{d}{=} ax + e$, where the equality is of probability distributions (i.e. $x$ and $ax+e$ have the same distribution). This unique stationary distribution is moreover power-law tailed with exponent $\mu$ given by $E(a^\mu) = 1$, as derived heuristically earlier (again, cf. the appendix). Thus $\{r_t\}$ converges to a power-law tailed distribution of exponent $\mu$.

**The (asymptotic) exponential random walk of absolute return**

Kesten's theorem, though technical in its details, rests actually on a classic result of the theory of random walks, which is therefore worth going through, as it signals a potential link between the theory developed here and the traditional random walk hypothesis. The starting point, similarly to the previous intuitive derivations, is to rewrite (35) as $\ln|r_t| = \ln|a_t r_{t-1}(1 + e_t/a_t r_{t-1})| = \ln|a_t| + \ln|r_{t-1}| + \ln|1 + e_t/a_t r_{t-1}|$, which simplifies when $|r| \to \infty$, and a fortiori if in addition $N \to \infty$, to

$$\ln|r_t| \sim \ln|r_{t-1}| + \ln|a_t|. \tag{38}$$

That is, the logarithm of absolute return follows asymptotically a random walk, with a mean increment $E[\ln(a_t)] < 0$ ($|a_t| = a_t$). Now, a random walk with negative mean increment is well-known to have, not a normal distribution, but an exponential one [58, sec. XII.5, 59, sec. 7.4, 38]. That is, $P(\ln|r| > x) \sim Ce^{-\mu x}$, where again $\mu$ is the solution to $E(a^\mu) = 1$, and $C$ is some constant; it follows that absolute return is asymptotically power-law tailed with exponent $\mu$, as intuitively derived earlier. In sum, it's neither price itself that follows a random walk, as Bachelier thought, nor its logarithm, as is dominantly assumed today, but, asymptotically, the logarithm of absolute return. A speculative price follows a sort of doubly multiplicative process.

The same observations apply to excess demand as well.

## 5 The human side of financial instability

Implicit in this conclusion, however, is the assumption that $|e_t/a_t r_{t-1}|$ is negligible when $|r_{t-1}| \to \infty$. This condition, which was already highlighted in the previous heuristic, is guaranteed to hold whenever, intuitively speaking, the exogenous term $e_t$ is milder than the feedback term $a_t r_{t-1}$, that is, when it has less probable extreme realizations. This is the intuitive meaning of the requirement that $E(|e_t|^\mu) < \infty$ in Kesten's theorem (see appendix, theorem 1, condition (iii)). In our particular case where even the external input is multiplicative, namely $e_t = a_t b_t$, this comes down to checking that $|b_t/r_{t-1}|$ is negligible asymptotically, that is, that the behavioral impulse is milder than return, which, as we said, is precisely the case so far, since, by the central limit theorem, $b_t$ is normal.

But the central limit theorem requires that expectations (of price changes and intrinsic values) are weakly correlated (at least among peers, namely among buyers and among sellers, for the correlation is negative between a buyer and a seller, as Bachelier observed, adding to the mildness of the behavioral impulse). A deep implication of Kesten's theorem is that, no matter how expectations are formed, no matter how dependent they are, and no matter how they are distributed, all the results so far are unaffected as long as $E(|e_t|^\mu) < \infty$, namely $E(|a_t b_t|^\mu) = E(a_t^\mu)E(|b_t|^\mu) = E(|b_t|^\mu) < \infty$; which holds in general.

A notable exception happens, however, when $b_t$ is itself power-law with exponent $\mu_b \leq \mu$ (in which case $b_t$ is wilder than $a_t$), for then $E(|e_t|^\mu) = E(|b_t|^\mu) = \infty$. Even then, however, return would still be power-law tailed, with the same exponent as the behavioral impulse ($\mu_r = \mu_b$), as a complementary to Kesten's theorem by Grincevićius implies [32]. In this case, the wildness of financial randomness would simply be a manifestation of the wildness of investors' expectations. More precisely, this theorem implies that

$$P(r > x) \sim P(b > x)E(a^{\mu_b})[1 - E(a^{\mu_b})]^{-1}, \tag{39}$$

given that here $e = ab$ (see appendix, theorem 3). Therefore, even in this case, the very trading activity, as summarized by $a = \alpha\beta N/L$, though it does no longer determine the shape of the distribution of return, plays nonetheless an important amplifying role, for it amplifies the likelihood of extreme events by $k''' = E(a^{\mu_b})[1 - E(a^{\mu_b})]^{-1}$. If by a wave of contagious over-optimism, for instance, which, as we shall see shortly, can indeed lead to a power-law collective impulse, investors rush to buy the security, the ensuing price change is likely to be extreme. But the greater the number $N$ of traders involved and the lower the liquidity $L$ in the market to meet this spontaneous mania, the more likely extreme will the price change be (possibly because these overheated traders would offer the highest prices to interest potential counterparts). Indeed, if $a$ is so high that $E(a^{\mu_b}) = 0.9$, e.g., then behavioral impulses beyond an extreme level $x$, which happen, say, merely once in a century, would induce equally extreme price changes that happen 9 times more likely, i.e., once every 9 years $(k''' = 0.9 \times 10 = 9)$. Even in this special case, in other words, the very logic of speculation compounds the wild psychosocial disturbances. But to what extent is a power-law tailed behavioral impulse mathematically plausible?

First of all, expectations have to be sufficiently *strongly interdependent* if this is to be the case; alternatively, expectations should be *extremely heterogeneous*, that is, in formal terms, their variance should be infinite. Otherwise, the behavioral impulse would remain normal, by the central limit theorem, no matter how expectations are individually determined[20]. Let's investigate these two possibilities.

## 5.1 Extremely divergent opinions

The empirical facts exclude this hypothesis directly. For assume the opinions $\theta_{it}$, for instance, are distributed across traders according to an infinite-variance power-law tailed distribution, while being independent; then by the generalized central limit theorem, $\bar{\theta}_t$ would converge (in a very active market) to a Levy random variable of index $\mu_\theta < 2$, which is asymptotically power-law with exponent $\mu_\theta$. But precisely in this case $\mu_r = \mu_b \leq \mu_\theta < 2$, at odds with the data. This, in essence, comes down to Mandelbrot's hypothesis, if, to mimic Fama's interpretation, we assume the news coming to the market place are perceived by traders in such extremely divergent ways as to induce readjustments in expectations that have infinite variance (across traders).

But we need not invoke a theorem, in the first place, to exclude this hypothesis; it is sufficient to recall that markets are more likely driven by mass psychology (and herd behavior), which tends to homogenize expectations. So let's therefore explore the more plausible alternative of strongly interdependent expectations.

## 5.2 Mimetic expectations

It is easy to show, still with Kesten's theorem, that mimetic expectations, as portrayed in Keynes's beauty contest metaphor, can naturally lead to a power-tailed average opinion $\bar{\theta}_t$ or $\bar{\phi}_t$ (opinions on price changes or intrinsic values), or average prediction error $\bar{\varepsilon}_t$ of the earlier formulation[21]. Keynes's point is that

*[...] professional investment may be likened to those newspaper competitions in which the competitors have to pick out the six prettiest faces from a hundred photographs, the prize being awarded to the competitor whose choice most nearly corresponds to the average preferences of the competitors as a whole; so that each competitor has to pick, not those faces which he himself finds prettiest, but those which he thinks likeliest to catch the fancy of the other competitors, all of whom are looking at the problem from the same point of view. It is not a case of choosing those which, to the best of one's judgment, are really the prettiest, nor even those which average opinion genuinely thinks the prettiest. We have reached the third degree where we devote our intelligences to anticipating what average*

*opinion expects the average opinion to be. And there are some, I believe, who practise the fourth, fifth and higher degrees.*

A trader must take into account the average prevailing opinion simply because whoever holds a view that goes against the market's overall sentiment, and acts accordingly, is sure to be ruined. André Orléan is a prominent defender of this 'self-referential hypothesis' today [55]. In George Soros's (philosophically inclined) writings, it appears under the name of reflexivity [60]. Formally, it comes down to assuming that a trader's opinion reflects the average opinion he expect to prevail in the market. What follows is a simple mathematical formulation of this intuition. We shall do so for the opinions about a rise in return, namely $\theta_i$, keeping in mind that the same formulation applies to the perceived value-price gaps $\phi_i$ as well, and to the prediction errors $\varepsilon_i$ alike.

Let $\theta_j^i$ denote trader $i$'s guess of trader $j$'s opinion about a change in return. It is natural to think of $\theta_i^i$ as $i$'s personal opinion prior to taking into account others'. The mimetic expectation hypothesis then simply says that trader $i$'s final opinion is

$$\theta_i = \sum_j w_{ij} \theta_j^i, \qquad (40)$$

where $w_{ij}$ is the weight $i$ assigns to $j$'s opinion (a measure of how important he thinks $j$'s opinion is); we have of course $\sum_j w_{ij} = 1$. There is no reason why these weights should stay fixed; on the contrary they can fluctuate as traders' perceptions of each other change. We assume they vary randomly. Equation (40) is better written as

$$\theta_i = w_{ii} \theta_i^i + \sum_{j \neq i} w_{ij} \theta_j^i, \qquad (41)$$

to highlight the fact that a trader's expectation is somewhere between his prior personal opinion and what he thinks others are thinking on average.

**A network of opinions**

The matrix $\mathbf{\Omega} = [\omega_{ij}]$, where $\omega_{ij} = w_{ij}$ if $i \neq j$ and $\omega_{ii} = 0$, defines a network among the traders, representing who values whose opinion (so is the matrix $[w_{ij}]$, of course; only we are avoiding 'self-edges'). We assume traders' expectations are strongly interdependent, not necessarily in the sense that anyone's opinion is directly relevant to anyone else (i.e. $\omega_{ij} > 0$, for all $i \neq j$, which would imply a 'complete network'), but in the sense that the network is *'strongly connected'*, that is, any pair of traders is connected at least indirectly, through other traders, or, in technical terms, there is at least one 'path' of some length $\lambda$ linking any pair of traders; for instance, even if trader $i$ has no reason a priori to take trader $j$'s opinion seriously ($\omega_{ij} = 0$), he may still end up taking it into account if there is some trader $k$ whose opinion he values ($\omega_{ik} > 0$) and who happens to value $j$'s opinion ($\omega_{kj} > 0$). As is well known, the existence of indirect links of length $\lambda$ corresponds to positive entries of the matrix $\mathbf{\Omega}^\lambda$ ($\sum_k \omega_{ik} \omega_{kj} > 0$, e.g., means there is at least one indirect link of length 2 from $i$ to $j$). Of course $\lambda = 1, 2,$ etc. correspond to the first, second, and higher degrees of guessing in Keynes's metaphor, while $\lambda = 0$ is the zeroth degree, namely the traders' prior personal opinions. In sum, a strongly connected network is such that, for any pair $(i, j)$, there is at least one positive $\lambda$ such that $[\mathbf{\Omega}^\lambda]_{ij} > 0$, or, equivalently, the matrix $\sum_{\lambda=1}^\infty \mathbf{\Omega}^\lambda > \mathbf{0}$ (i.e., its entries are all positive); $\mathbf{\Omega}$ is then said to be 'irreducible'. Equivalently, a strongly connected network is such that the matrix

$$\mathbf{K} = \sum_{\lambda=0}^\infty \mathbf{\Omega}^\lambda > \mathbf{I}, \qquad (42)$$

where $\mathbf{I}$ denotes the identity matrix of size $N$. This multidimensional multiplier amplifies any vector it applies to, like the one-dimensional multiplier $k$. Its entries may diverge to infinity, when the series diverges. This is not the case here, however. Indeed all row-sums in the matrix $\mathbf{\Omega}$ are below unity by construction ($\sum_j \omega_{ij} \leq 1$), and at least one row-sum can be assumed to be strictly below unity, for at least one trader can be assumed to value his prior judgment ($w_{ii} > 0$). It follows that $\mathbf{\Omega}$ has a spectral radius $\rho(\mathbf{\Omega}) < 1$ (spectral radius meaning the maximum absolute eigenvalue) [22]. Therefore the series in (42) converges and $\mathbf{K} = (\mathbf{I} - \mathbf{\Omega})^{-1}$, and is thus finite. (This mirrors the so-called Leontief inverse.) Thus, a strongly connected network of opinions with $\rho(\mathbf{\Omega}) < 1$ is such that anyone's opinion will eventually matter for anyone else, whether directly or indirectly; only an indirect distant influence vanish exponentially with its lengths $\lambda$, which is reasonable.

**Wildly amplified opinions**

Let $\pi_i = w_{ii}\theta_i^i$ denote the weighted prior personal opinions, and let $\boldsymbol{\pi} = [\pi_i]$ be the corresponding vector; let $\boldsymbol{\theta} = [\theta_i]$ be the vector of the final individual opinions. Now, only by coincidence would an individual's guess of others' guesses be accurate; thus we should consider the misrepresentations of opinions $m_j^i = \theta_j^i - \theta_j$, and their weighted averages $m_i = \sum_j \omega_{ij} m_j^i$; let $\mathbf{m} = [m_i]$ and $\boldsymbol{\varepsilon} = \boldsymbol{\pi} + \mathbf{m}$. Then (41) becomes

$$\boldsymbol{\theta} = \mathbf{\Omega}\boldsymbol{\theta} + \boldsymbol{\varepsilon}, \tag{43}$$

implying $\boldsymbol{\theta} = \mathbf{K}\boldsymbol{\varepsilon}$. By this social feedback, therefore, both the prior personal opinions and the misrepresentations of others' opinions are amplified. Of course, (43) is but the multidimensional version of the model we used earlier. Actually, Kesten's original theorem is about the multidimensional case, while Vervaat, Goldie, and others focused on the one-dimensional case, while extending Kesten's assumptions.

But, as previously, it is more realistic to reinterpret the instantaneous feedback in (43) as the stationary limit of the guessing process. Thus assume $\theta_{jt}^i = \theta_{jt-1} + \nu_{jt}^i$; that is, assume $i$ expects a variation of $\nu_{jt}^i$ from $j$'s previous opinion, assuming he can observe this previous opinion. Then we should consider time-varying weights $\omega_{ijt}$ and their corresponding matrices $\mathbf{\Omega}_t = [\omega_{ijt}]$. Letting $\nu_{it} = \sum_j \omega_{ijt}\nu_{jt}^i$, $\mathbf{v}_t = [v_{it}]$, and $\boldsymbol{\omega}_t = \boldsymbol{\pi}_t + \mathbf{v}_t$, (41) becomes

$$\boldsymbol{\theta}_t = \mathbf{\Omega}_t\boldsymbol{\theta}_{t-1} + \boldsymbol{\omega}_t, \tag{44}$$

namely, a multidimensional stochastic recurrence equation. It has a unique stationary solution $\boldsymbol{\theta}$ under essentially the same conditions as in the one-dimensional case, notably $E(\|\mathbf{\Omega}_t\|^\delta) < 1$, for some $\delta > 0$, where the norm is the Euclidean one (this is of course but an extension of $E(|a_t|) < 1$, or $E(\ln|a_t|) < 0$, more generally).

Moreover, as in the one-dimensional case, the mimetic guessing process leads to power-law tailed individual opinions $\theta_{it}$, with the same tail exponent $\mu_\theta$ given by

$$\lim_{t \to \infty}(E\|\mathbf{\Omega}_1 \cdots \mathbf{\Omega}_t\|^{\mu_\theta})^{\frac{1}{t}} = 1, \tag{45}$$

which is the multidimensional equivalent of $E(|a_t|^\mu) = 1$. More generally, Kesten's theorem says that all linear combinations $\mathbf{c}^T\boldsymbol{\theta}$ of the stationary solution $\boldsymbol{\theta}$, for any nonzero vector $\mathbf{c}$, are power-law tailed with exponent $\mu_\theta$; that is, $P(\mathbf{c}^T\boldsymbol{\theta} > x) \sim \Psi(\mathbf{c})x^{-\mu_\theta}$, where $\Psi$ is some positive function [31, 34, 39]. It follows that $\bar{\theta}_t$, and thus $b_t$, is power-law tailed with exponent $\mu_\theta$ (taking $\mathbf{c}^T = N^{-1}[1,...,1]$). Therefore, mimetic expectations could indeed explain, a priori, the power-law tails of return. In this case, the tail exponent would be entirely determined by the weights traders assign to each other's opinions.

But, while theoretically plausible, this hypothesis raises various problems, starting from the very applicability of Kesten's theorem. First, the opinion weights can hardly be assumed to vary randomly all the time, at the rhythm of the high-frequency market fluctuations (the empirical power laws hold even at this low time scale). Second, even if they did, they would hardly do so in an iid (or stationary and ergodic) way, as Kesten's theorem requires. Also, similarly to the previous hypothesis of extremely divergent opinions, the empirical fact, at least for stock returns ($\mu \approx 3$), may invalidate this one also. Indeed, in a very active market ($N \to \infty$), the steady state average opinion $\bar{\theta}$, which results from the mimetic guessing process, being the sum of a large number of power-law tailed variables, could converge to a stable variable, namely either a Gaussian (if $\mu_\theta \geq 2$) or a Levy variable (if $0 < \mu_\theta < 2$), at odds with the data. Only, this conclusion assumes that the individual opinions, though initially interdependent by construction, will eventually, at the steady state, become weakly dependent enough[23]. Finally, why would subjective opinion weights vary in such a precise way as to induce an average opinion consistently and almost universally distributed as $x^{-3}$? In contrast, given the relative invariance and similarity of the basic trading practices worldwide, it is not unreasonable to assume that the distribution of trading, namely the number of trades and liquidity, is overall time-invariant and almost universal. So the previous explanation remains the more plausible one.

## 6 Correlations between assets

It remains to address a final hypothesis: that the power laws emerge by the correlations between assets, as some prices tend to move together. The theory wouldn't be general if it ignores this issue. Fortunately, the previous model of interdependent expectations applies isomorphically to this hypothesis also; so Kesten's theorem is still all that is needed! The key concept here is the change $a_{ij}$ in $r_i$, the return of an asset $i$, caused by a unit change in $r_j$, the return of an asset $j$ (ceteris paribus). These coefficients are not fixed; we assume they fluctuate randomly. Naturally, the effect that $r_i$ has upon itself, namely its feedback, is given by $a_{ii} = a_i = \alpha_i \beta_i N_i / L_i$, as we know from the main model; we need also consider the inputs $e_i = a_i b_i$. In sum, by this interdependence, returns are given by

$$r_i = \sum_j a_{ij} r_j + e_i. \tag{46}$$

Letting $\mathbf{r} = [r_i]$, $\mathbf{A} = [a_{ij}]$, and $\mathbf{e} = [e_i]$, this reduces to

$$\mathbf{r} = \mathbf{A}\mathbf{r} + \mathbf{e}, \text{ or } \mathbf{r}_t = \mathbf{A}_t \mathbf{r}_{t-1} + \mathbf{e}_t, \tag{47}$$

depending on whether the cross-effects and the feedbacks happen instantaneously or after a one period lag. Thus, the analysis of the mimetic expectations model can be applied here also. If the conditions of Kesten's theorem are met, which could be decided only empirically, returns would be power-law tailed, by the interdependence among the assets, more generally, rather than the mere feedbacks. Then the network of interdependences among the assets, or the matrix $\mathbf{A}$, would be of great theoretical importance.

But, at the same time, this hypothesis suffers the same problems as the mimetic expectations one, notably the requirement that the links among assets, or $\{\mathbf{A}_t\}$, should evolve in an iid (or stationary and ergodic) way. Most likely, there is greater persistence in $\{\mathbf{A}_t\}$ (which needs not be stationary ergodic). Moreover, because assets of different kinds can affect one another (e.g. a stock, a bond and a currency), this hypothesis would predict that their returns are power-law, with the same exponent; this is hardly the case empirically.

So these theoretical refinements of the basic model of speculative feedback are not likely to add anything substantial to it. Of course, this calls for an empirical check, ultimately. In this respect also, both the mimetic expectations and the interdependent assets hypotheses present an immediate practical problem: the opinion weights and the assets' cross-effects are not recorded data, and could be hardly so. Indeed, as we said, the very idea of constantly fluctuating opinion weights is not a realistic one. As for the cross-effects, if traditional regression analysis can be performed to estimate them, the results may be unreliable, since they would involve wildly random variables (whose variances and covariances can be very high). The basic model, in contrast, suffers no such problems.

# 7 Conclusion

In sum, the wildness of financial instability is intrinsic to the very speculative nature of financial activity, as a result of its endogenous amplifying nature. At any instant, the extent of this potentially destabilizing power depends on the ratio of the number of orders coming to the market and the market's capacity of absorbing them, namely its liquidity. Bad news can cause crises, but only to the extent that they unfold a speculative chain reaction of distress sell orders, particularly in periods of low liquidity. Investors' psychology can be decisive as well, but also as a mere trigger. Individual psychological impulses per se can hardly account for the wildness of financial markets, as they tend to cancel overall. When they become strongly interdependent, however, notably by the herd behavior of investors, they can cause greater instability: herd behavior can provoke a collective panic, which leads to crises, by the vicious circle of speculation. In this case, a purely psychosocial feedback adds to the mechanical feedback of the speculative trading itself. Correlations between assets can also add to the instability, by spreading local disturbances to the whole economy, and beyond. Sill, speculative feedback remains the main cause of financial fluctuations, as it explains, in the most natural way, their power-law nature.

---

[1] A review and synthesis of Bachelier's work with those of Osborne, Cootner, Samuelson and Mandelbrot, notably [6-10].

[2] Because a random walk explores negative values, a price movement cannot be a random walk (for a negative price is not meaningful). A natural alternative, which has become standard, is to assume the logarithm of price instead as following a random walk, in which case price $\equiv$ exp[ln(price)] is guaranteed to be positive. Price is then said to follow an 'exponential random walk' (or geometric random walk). Besides positivity, however, this formulation is also more convenient because price formation is more likely a multiplicative than an additive process. But, as this paper will conclude, it is actually the logarithm of (absolute) return that follows a random walk (asymptotically). The price process is therefore doubly multiplicative.

[3] Cf. Bouchaud's introduction to this view [29].

[4] We take the purely psychosocial dimension of speculation as given, as an exogenous factor, in the same way that we treat the announcements of real news. In this regard, the behavioral literature also comes down to the old paradigm of exogenously induced price fluctuations, to the extent that it resorts to human social psychology (which an economist takes as given), rather than the mechanical functioning of markets. But, on the other hand, the social psychology of trading can equally be seen as an endogenous dimension, for psychosocial stimuli are not input from outside the market. So we should keep in mind the potential ambiguity of this endogenous-exogenous dichotomy. The ambiguity is compounded, moreover, given the following trivial fact: what ultimately affect prices is less the news themselves, in their objectivity, than the subjective feelings they impress on investors (put differently, news affect prices only insofar as they are processed by human brains). This is why we shall rarely separate the effects of news from purely psychological stimuli, in a clear-cut way. Still, the convention in this paper is to call exogenous any mechanism

other than supply and demand itself, that is, the trading activity in its mechanical unfolding, and to treat all mechanisms that act as triggers, namely real news and psychosocial stimuli, as exogenous factors.

[5] There are earlier attempts to model price changes as random stochastic growth model. ARCH and GARCH models, for instance, though not originally thought as such, come down to a multiplicative random growth process. Only, as often reproached, these are ad hoc (postulated, not derived) models, which hardly correspond to a clearly articulated underlying economic process. Also, they place no restriction a priori on the tail exponent of returns, and says nothing about other power-law distributed variables, such as volume. In contrast, Lux and Sornette's reinterpretation of the dynamics of 'rational bubbles' in terms of a multiplicative random growth is more theoretically grounded [30]. Thus, beforehand, this could explain the power-law tails of the distribution of returns; only, as they quickly pointed out, the very idea of a 'rational bubble' restrict the power-law exponent below unity, at odds with the data. Cf. Lux's review for similar works [28].

[6] In a similar vein, Delong et al. highlight the destabilizing role of what they call 'positive feedback traders', namely traders who buy when they observe price rising and sell when they fall [13]. Feedback has a more general meaning here, however.

[7] In particular we shall avoid statements like 'prices "fully reflect" all available information', a proposition, which, as Fama conceded, 'is so general that it has no empirically testable implications' [4]. Also, when interpreted literally, it is a self-defeating proposition, even from a utility-maximization 'rational expectations' perspective, as Grossman and Stiglitz have famously pointed out [45]. But we shall also avoid 'utility maximization' altogether, which is also well-known to have no deep observable implication for collective behavior in general (the Sonnenschein-Mantel-Debreu theorem), forcing most authors to reduce the entire economy to one 'infinitely-lived representative agent', to avoid the aggregation problem!

[8] Equation (3), and similar equations to come, can be viewed as a first order Taylor approximation of a possibly more general (nonlinear) tendency, which is justified by the fact that $r_t^e$ is arguably small.

[9] Implicit in the efficient market hypothesis is the belief that investors' expectations converge, on average, to the actual values, if they use every relevant piece of information in forming these expectations. Unfortunately, as we said, this intuition is popularly expressed in the abstract language of 'rational expectations', which makes a series of unrealistic assumptions about human capacity for prediction (notably, that people form their expectations as if they were using a correct model of the entire economy, and that their collective expectations reduce to that of an 'infinitely-lived representative agent'). Also, the subjective expectations are assumed to behave as mathematical expectations, conditioned on all available information, an assumption that is justified by its corresponding to the best use of the available information, as is well-known from elementary statistics (best in the sense of minimum mean squared error of prediction). But in the financial world of wild randomness, where the variance (or even the mean) of many variables is so large (in some cases even infinite), this is not a useful prediction recipe (the minimized mean squared error can still be as large as to make the predicted value of no practical use). Thus we shall not use this abstract mainstream approach to expectations. Rather we model them in a simpler and more general way throughout.

[10] Because in reality $N$ is also a random variable, this and all following statements about limit behaviors are more technical then it appears at first glance. For extensions of the law of large numbers and the central limit theorem to a sum of a random number of terms, see Embrechts et al.'s *Modelling Extremal Events for Insurance and Finance*, sec. 2.5 [48]. But throughout we shall keep the technicality to a minimum.

[11] The paradox was already apparent earlier, in (15), concerning the bubble's growth rate $\rho$, which increases with $\gamma$ and decreases with $\alpha$; that is, the more investors are focused on fundamentals, the faster the bubble develops, and the more speculative they are, the slower it develops!

[12] A useful terminology: if $X$ is power-law tailed, we say it is *wilder* than $Y$ (or that $Y$ is *milder* than $X$) if $Y$ is normal, exponential, or any variable whose moments are finite at all orders. If both $X$ and $Y$ are power-law tailed with exponents $\mu_x$ and $\mu_y$, resp., we say that $X$ is wilder than $Y$ (or $Y$ is milder than $X$) if $\mu_x < \mu_y$.

[13] For more discussion on this, see Farmer's 'Market force, ecology, and evolution' [47].

[14] Kyle's classic theory also predicts a linear price impact [51].

[15] More precisely, Cont et al. established a good linear fit for $E(\Delta P \mid q)$, where $\Delta P$ is the mid-price change expressed in number of ticks; but since $P$ vary more slowly than $\Delta P$, particularly over short time periods, we consider this finding to imply also a good linear fit for $E(r \mid q) \approx P^{-1} E(\Delta P \mid q)$.

[16] We thank J.-P. Bouchaud for pointing out a different view on the 'square root law' in the literature.

[17] We could have defined the prediction error as $\varepsilon_{it} = r_t^e - r_t$ from the beginning; only in this case we would have faced an arbitrary minus sign in equilibrium return, in the random walk model.

[18] Indeed when $N$ is power-law tailed, the ratio $N/L$ is also power-law tailed, and its exponent is the same as $N$, except when $L^{-1}$ is wilder than $N$, in which case the ratio inherits the exponent of $L^{-1}$ (cf. the generalities on power laws in Gabaix's review [56]). So if $L^{-1}$ is uniform, Gaussian, exponential, or any other relatively mild variable, the ratio is power-law tailed with the same exponent as $N$. Now, if $L$ takes values spread near zero, it is easy to show that $L^{-1}$ would be power-law tailed with exponent 1 (cf. Newman's review, sec. IV.B [57]). This would imply that $a = \alpha\beta NL^{-1}$ is power-law tailed with exponent 1. But we exclude this case, because only exceptionally would liquidity be close to drying completely.

[19] Also, we need not assume that there is an objective 'intrinsic value' $F$ for any security, whose value would be identically assessed by investors, moreover; instead all we need to assume are the subjective guesses $F_i$, namely investors' opinions about the security's worth, given the information available to each of them and their respective capacities of processing it, whether these guesses match an objective economic reality or not. Cf., on this issue, Orléan's critique of the objectivity of financial value [55, 56].

[20] Diversity and interactions being two important dimensions of complexity, the following discussion has a much broader and more fundamental scope than this particular application, in that it highlights their natural link with power laws, which are a key signature of complexity.

[21] The so-called Ising model in physics has also been applied to herd behavior, as it may exhibit power-law average opinions, though not in a generic way. For an introduction to this literature see chapter 20 of Bouchaud and Potters' *Theory of Financial Risk and Derivative Pricing,* and the references therein [22].

[22] Cf., e.g., Meyer's *Matrix analysis*, exercice 8.3.7, p. 685 [61].

[23] Weak temporal dependence, however, is sure to emerge in these multiplicative random processes, in the technical sense of 'strong mixing' at geometric rate, which is, intuitively speaking, the property of certain stationary ergodic processes of exhibiting, at an exponential rate, a near-independence between two remote states [39, 42]. This is therefore a natural explanation of the unpredictability of returns.

## Appendix: Kesten's theorem and its extensions

**Theorem 1** (Kesten, Vervaat, Goldie [31, 33, 36]). Consider the random equation (∗) $x \stackrel{d}{=} ax + e$ (an equality in distribution), where the unknown $x$ is a random variable and $(a,b)$ is a pair of random variables independent of $x$. Assume there is a $\mu > 0$ such that (i) $E(|a|^\mu) = 1$, (ii) $E[|a|^\mu \max(\ln(|a|), 0)] < \infty$, and (iii) $E(|e|^\mu) < \infty$; assume also that (iv) $\ln|a|$, given $a \neq 0$, is nonarithmetic (i.e., its realizations are not just integer multiples of a given real number). Then (∗) has a solution $r$ unique in distribution. This distribution is power-law tailed; i.e., when $x \to \infty$, $P(r > x) \sim C_+ x^{-\mu}$ and $P(r < -x) \sim C_- x^{-\mu}$, where $C_+$ and $C_-$ are nonnegative constants; or, more compactly, $P(|r| > x) \sim C x^{-\mu}$, $C = C_+ + C_-$. If (iv) $P(a \geq 0) = 1$, $C_+ \neq C_-$; (v) otherwise, $C_+ = C_-$. Also, $C = C_+ + C_- \neq 0$ except for the trivial case where (vi) $(1-a)^{-1} e$ reduces to a (nonrandom) constant. Finally, the stochastic process $\{r_t\}$ generated by the recurrence $r_t = a_t r_{t-1} + e_t$, where $(a_t, b_t)$ are (vii) independent copies of $(a,b)$, converges in distribution to $r$ as $t \to \infty$, no matter how initiated; i.e., $r_t$ converges to a (stationary) power-law distribution, with exponent $\mu$.

*Remarks*. Originally, Kesten's theorem was formulated in a multidimensional setting, i.e. with $N$-dimensional vectors $r_t$ and $e_t$ and $N \times N$ nonnegative invertible matrices $a_t$; it is extended to wider classes of matrices by many authors, notably LePage et al. [34, 40].

**Theorem 2** (Kesten, Vervaat, Brandt [31, 33, 35]). Consider the stochastic recurrence equation (∗) $r_t = a r_{t-1} + e_t$, where $\{(a_t, e_t)\}$ is a stationary and ergodic sequence. If (i) $-\infty \leq E(\ln|a_t|) < 0$ and (ii) $E[\max(\ln|e_t|, 0)] < \infty$, or if (iii) $P(a_t = 0) > 0$, then (∗) has a unique stationary solution. Also, for any arbitrary random initial variable $r_0$, the process $\{r_t\}$ converges in distribution to this solution.

*Remarks*. Condition (i) is implied by conditions (i), (ii) and (iv) of theorem 1 [36]. In its original form by Vervaat, and to some extent by Kesten, this theorem requires an iid sequence $\{(a_t, e_t)\}$; the extension to a stationary and ergodic sequence is by Brandt. When $\{a_t\}$ is irreducible, aperiodic, stationary, finite state space Markov chain, theorem 1 also holds, as shown by de Saporta [43].

**Theorem 3** (Grincevićius, Grey [32, 37]). Consider a random pair $(a,e)$ where: (i) $e$ is power-law tailed with exponent $\mu_e$, i.e. $P(e > x) \sim C x^{-\mu_e}$; (ii) $P(a \geq 0) = 1$; (iii) $E(a^{\mu_e}) < 1$; (iv) $E(a^\delta) < \infty$ for some $\delta > \mu_e$; and (iv) $E[\max(\ln(|e|), 0] < \infty$. Then there is exactly one distribution for $r$ such that $r \stackrel{d}{=} ar + e$, which is moreover power-law tailed with the same exponent $\mu_e$ as $e$; more precisely, $P(r > x)/P(e > x) \to [1 - E(a^{\mu_e})]^{-1}$, as $x \to \infty$. Also, the process $\{r_t\}$ generated by the recurrence $r_t = a_t r_{t-1} + e_t$, where $(a_t, e_t)$ are independent copies of $(a,e)$, converges in distribution to $r$, as $t \to \infty$, and thus has a limit distribution which is power-law tailed with exponent $\mu_e$.